\documentclass[twocolumn,showpacs,amsmath,amssymb,prl,aps,superscriptaddress, floatfix]{revtex4}

\usepackage{color}
\usepackage{graphicx}
\usepackage{xspace}

\begin{document}

\title{Vortex core switching by coherent excitation  with single in-plane magnetic field pulses.}

\author{Markus Weigand}
\email[]{mweigand@mf.mpg.de}
\affiliation{Max-Planck-Institut f\"ur Metallforschung, 70569 Stuttgart, Germany}

\author{Bartel Van Waeyenberge}
\email[]{Bartel.VanWaeyenberge@UGent.be}
\affiliation{Max-Planck-Institut f\"ur Metallforschung, 70569 Stuttgart, Germany}
\affiliation{Department of Subatomic and Radiation Physics, Ghent University, 9000 Gent, Belgium}

\author{Arne Vansteenkiste}
\affiliation{Department of Subatomic and Radiation Physics, Ghent University, 9000 Gent, Belgium}

\author{Michael Curcic}
\affiliation{Max-Planck-Institut f\"ur Metallforschung, 70569 Stuttgart, Germany}

\author{Vitalij Sackmann}
\affiliation{Max-Planck-Institut f\"ur Metallforschung, 70569 Stuttgart, Germany}

\author{Hermann Stoll}
\email[]{stoll@mf.mpg.de}
\affiliation{Max-Planck-Institut f\"ur Metallforschung, 70569 Stuttgart, Germany}

\author{Tolek Tyliszczak}
\affiliation{Advanced Light Source, LBNL, 94720 Berkeley, CA, USA}

\author{Konstantine Kaznatcheev}
\affiliation{Canadian Light Source, 94720 Saskatoon, S7N 0X4, SK, Canada}

\author{Drew Bertwistle}
\affiliation{Canadian Light Source, 94720 Saskatoon, S7N 0X4, SK, Canada}

\author{Georg Woltersdorf}
\affiliation{Institut f\"ur Experimentelle und Angewandte Physik, Universit\"at Regensburg, 93040 Regensburg, Germany.}

\author{Christian H. Back}
\affiliation{Institut f\"ur Experimentelle und Angewandte Physik, Universit\"at Regensburg, 93040 Regensburg, Germany.}

\author{Gisela Sch\"utz}
\affiliation{Max-Planck-Institut f\"ur Metallforschung, 70569 Stuttgart, Germany}

\date{\today}

\begin{abstract}

The response of magnetic vortex cores to sub-ns in-plane magnetic field pulses was studied by time-resolved X-ray microscopy. Vortex core reversal was observed and the switching events were located in space and time. This revealed a mechanism of coherent excitation by the leading and trailing edges of the pulse, lowering the field amplitude required for switching. The mechanism was confirmed by micromagnetic simulations and can be understood in terms of gyration around the vortex equilibrium positions, displaced by the applied field.

\end{abstract}


\maketitle

The magnetic vortex is a typical ground state configuration of micron and sub-micron sized ferromagnetic thin film structures \cite{Raabe2000}. It minimizes the stray field energy by forming an in-plane curling magnetization. In order to avoid a singularity in the center of the structure, the magnetization turns out-of-plane, forming the vortex core which can point either up (vortex polarization $p=+1$) or down ($p=-1$). This configuration is very stable; static out-of-plane magnetic fields of about 0.5 Tesla are required to switch the core polarization \cite{Okuno2002}. The vortex also has a specific excitation mode, the so-called gyrotropic mode, which can be excited by an oscillating in-plane magnetic field \cite{Huber1982, Argyle1984}. It corresponds to a circular motion of the vortex around its equilibrium position. 

Recently, it was discovered that switching the polarization of the vortex core is not only possible by static fields, but also by excitation of the gyration mode \cite{Vanwaeyenberge2006}. In this case, only field strengths of a few millitesla are needed. Micromagnetic modeling of these experiments revealed that the vortex switching occurs by the creation and subsequent annihilation of a vortex-antivortex pair \cite{Vanwaeyenberge2006}. This discovery has triggered a variety of studies on vortex core switching by various excitation methods. In particular, it was found by micromagnetic simulations that the polarization may also be switched by very short in-plane magnetic field pulses. It was shown that switching times as short as 40\,ps are possible \cite{Hertel2007}. Except for a brief report on switching events observed with 2\,ns long spin polarized currents pulses \cite{Yamada2008}, the possibilities of such excitations have not been explored experimentally so far.  

In this work, we have experimentally investigated vortex core switching by  in-plane magnetic field pulses, using time resolved magnetic X-ray microscopy. By taking advantage of the pulsed nature of synchrotron light, stroboscopic imaging was set-up at Scanning Transmission X-ray Microscopes  (STXM) \cite{Kilcoyne2003} at beamline 11.0.2 of the Advanced Light Source and 10ID-1 of the Canadian Light Source. Using the X-ray Magnetic Circular Dichroism (XMCD) \cite{Schutz1987} as a contrast mechanism, magnetization dynamics could be studied with 25\,nm spatial and 70\,ps temporal resolution.

The samples studied in this work  are 500\,nm\,$\times$\,500\,nm and 1\,$\mu$m\,$\times$\,1\,$\mu$m square-shaped magnetic Permalloy (Ni$_{80}$Fe$_{20}$) elements with a thickness of 50\,nm. These structures are defined on top of a 2.5\,$\mu$m wide, 150\,nm thick Cu stripline, using e-beam lithography, thermal evaporation and lift-off processes.  

A fast digital pulse generator, synchronized with the synchrotron light flashes was used to send rectangular current pulses across the stripline, which generates in-plane magnetic field pulses.  The amplitude and length of the pulses was accurately controlled and the rise and fall time were about 180\,ps. The field strengths are calculated from the currents and have an estimated systematic error of about 15\%. The structure was pumped with repetition frequencies between 20 and 40\,MHz for billions of times in order to record one series of time resolved images.  

In this way, we studied the response of the vortex core to magnetic field pulses with a length of 700\,ps.  Below a field strength of about 12\,mT the vortex is observed to be displaced by the pulse and excited into a slowly damped gyrotropic motion (see row (b) in Fig.\,\ref{fig1} and Movie 1 of the supplementary material \footnote{\label{EPAPS}See EPAPS Document No. [number will be inserted by publisher]}). For higher field strengths, a second vortex of opposite polarity appears. Both vortices gyrate with opposite directions (see row (c) of Fig.\,\ref{fig1} and Movie 2 of the supplementary material\ref{EPAPS}). 
This behavior can be explained by the nature of our stroboscopic technique:  If the amplitude is sufficiently strong to switch the vortex core with each pulse, we expect to record a superposition of both the up and the down vortex responses in each image. This is illustrated in row (d) of Fig.\,\ref{fig1} and Movie 2 \ref{EPAPS}, where simulations \footnote{\label{sims}Micromagnetic simulations where performed with the OOMMF code\,\cite{Donahue1999} and parameters replicating the experimental conditions: saturation magnetization $M_s=736\times10^3$\,A/m\,\cite{Vansteenkiste2008}, exchange stiffness $A=13\times10^{-12}$ J/m, the gyromagnetic ratio $\gamma=2.21\times10^{5} \frac{m}{A\dot s}$, crystalline anisotropy constant $K_1=0$, cell size in $x$- and $y$-direction was 5\,nm, in the $z$-direction 50\,nm. The damping parameter $\alpha$ was adjusted to fit the decay of the post-switching gyration amplitude in the different samples ($\alpha=0.008$ for Fig.\,\ref{fig1} and \ref{fig2} and $\alpha=0.03$ for Fig.\,\ref{fig3}).} of the out of plane magnetization of these two events were added and convoluted with the experimental resolution \footnote{\label{resolution}The experimental resolution was about 70\,ps in the time domain (given by the synchrotron bunch length) and about 30\,nm in the spatial domain (given by the zone plate).}.

In these superimposed images, we observe that the vortex core is accelerated by the leading edge of the magnetic field pulse and moves smoothly away from the center of the element during the first 750\,ps of the pulse.  At a delay of 1\,ns  (just after the trailing edge of the pulse), multiple small contributions to the magnetic contrast can be seen. Looking at the corresponding simulations in Fig.\,\ref{fig1} row (d), this can be explained by two reasons: First, the vortex-antivortex formation associated with the core reversal leads to additional magnetic contrast. Second, the two images of the vortex core nearly overlap, hence the magnetic contrast cancels partially. The asymmetric contrast in the experiment can be explained by small asymmetries between the up and down states of the vortex core \cite{Chou2007,Curcic2008}.  After 1.5\,ns, the two images of the vortex can be seen in an undisturbed, damped gyration. 
Assuming that the vortex switching occurs after the end of the excitation pulse (frame 5 in Fig.\,\ref{fig1}), the trajectories of both vortex states with their switching events (up-to-down and down-to-up) were separated and compared to the simulations\ref{sims} shown in Fig.\,\ref{fig2}.  The remarkable agreement between simulation and experiment, especially the position of the switching and the drop of gyration amplitude following the switching clearly support the initial assumption that the switching occurs at the end of the excitation pulse.

\begin{widetext} 

\begin{figure}[!htb]
\centering
\includegraphics[width=\linewidth]{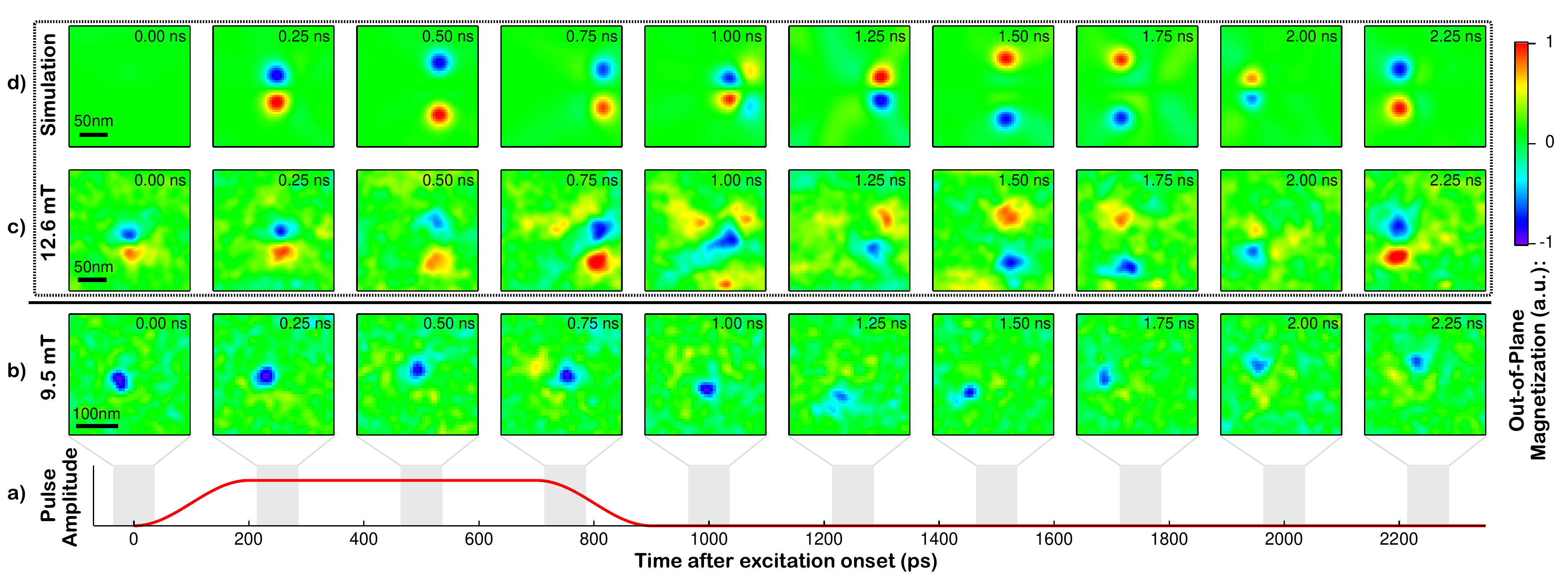}
\caption{\label{fig1} 
The out-of-plane magnetization of the center section of 500\,nm large, square permalloy element excited by in-plane field pulses (shown in row (a)) with a length of 700\,ps  is displayed in time steps of 250\,ps. Row (b) and (c) show experimental measurements of the reaction of the vortex core excited at 9.5 and 12.6\,mT, respectively. For comparison with row (c), row (d) depicts the superposition of the simulated out-of-plane magnetization of  vortex core up-down and down-up switching, convoluted with the experimental resolution\ref{resolution}.}
\end{figure}
\end{widetext}

\begin{figure}
\centering
\includegraphics[width=\linewidth]{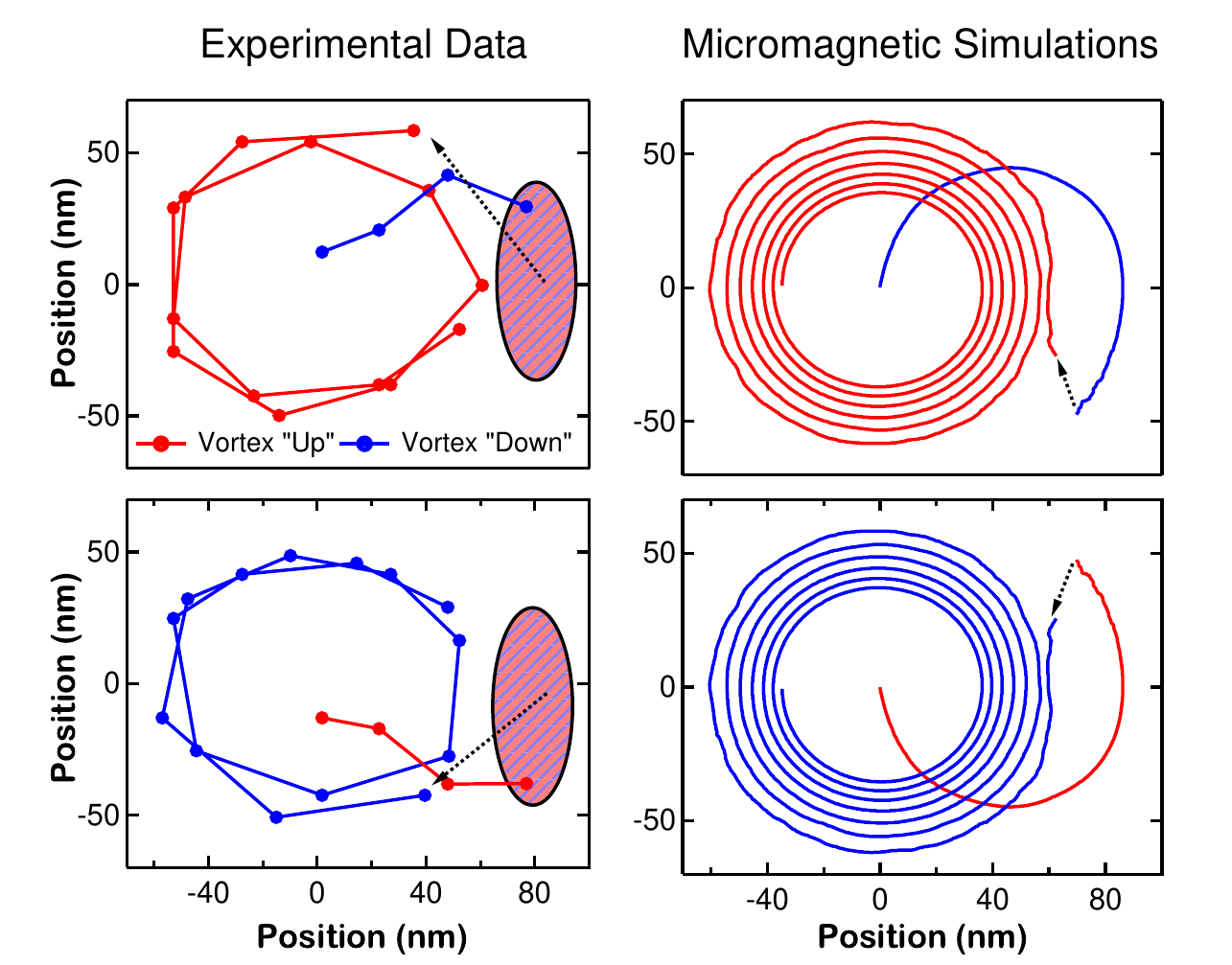}
\caption{\label{fig2} The left panels show separate plots of the trajectories of both the up and down vortex, as visible in the frames of Fig.\,\ref{fig1} row (c). The hatched area replaces the vortex position at the assumed switching time, where it could not be accurately tracked. In comparison, the right panels show the corresponding micromagnetic simulation using parameters derived from the observed sample\ref{sims}, under identical excitation, with an arrow connecting the positions of the vortex before and after switching. }
\end{figure}

\begin{figure}
\centering
\includegraphics[width=\linewidth]{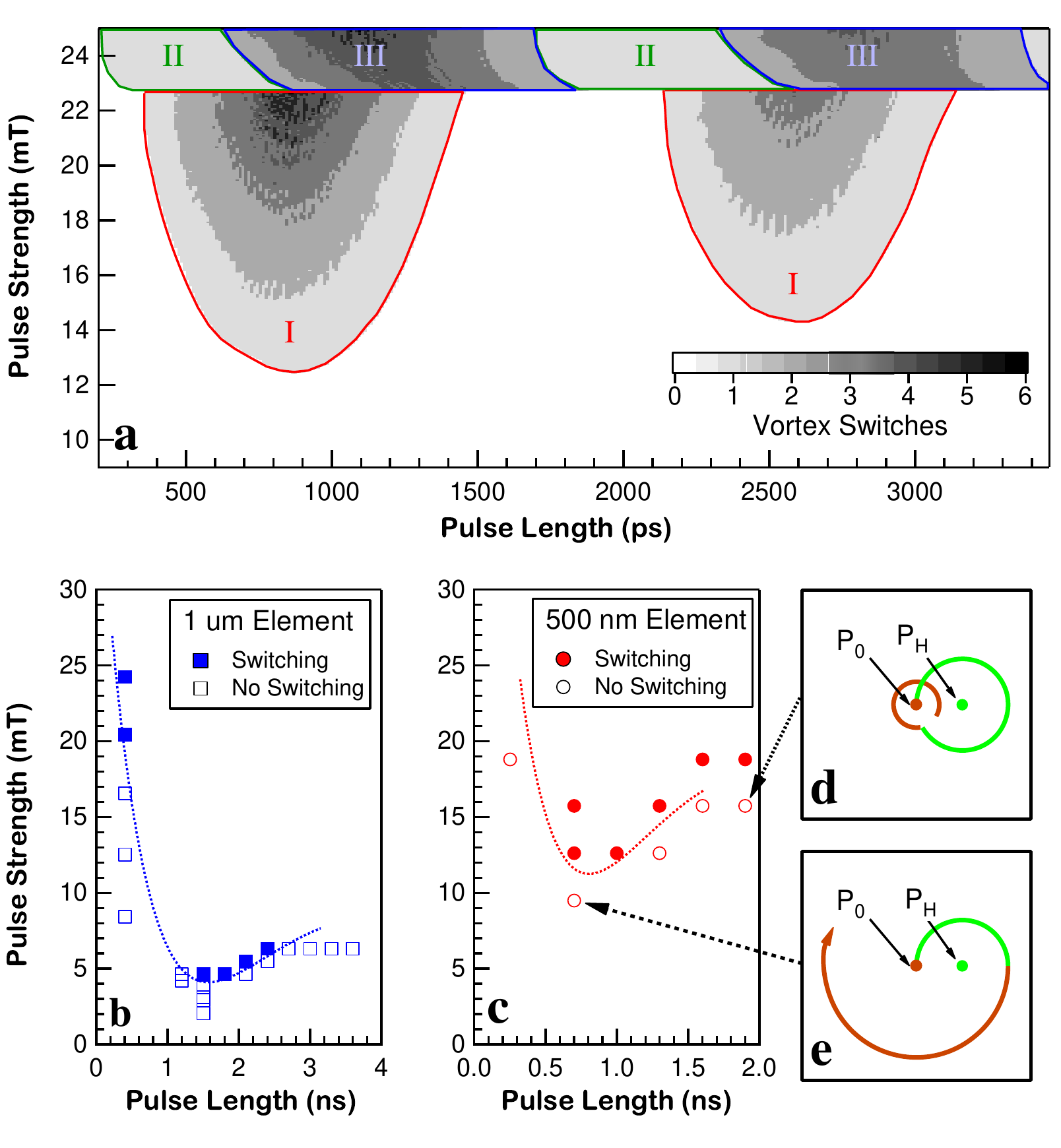}
\caption{\label{fig3} Panel a):  Diagram of simulated core switching events vs. pulse length and strength for 500\,nm structures. There are 3 prominent regions marked: I (red): Switching occurring after the trailing edge of the pulse. II (green): Pulse length independent switching at the leading edge of the pulse. III (blue): Leading edge switching followed by additional switching at the trailing edge of the pulse. 
Panels b) and c): Experimental results for 1\,$\mu$m (b) and 500\,nm (c) elements respectively. The solid dots indicate the parameter combinations where the vortex core is switching, the rings where it is not switching. Panels d) and e) illustrate the vortex core motion during (green) and after (brown) the pulse around the respective gyration centers $P_F$ and $P_0$.}
\end{figure}

In the experiment on the 500\,nm sized elements, an increase of the pulse length to 1300\,ps or a decrease  to 300\,ps with the same amplitude (12.6\,mT) suppresses the  switching, illustrating the importance of the pulse length for core reversal.

To investigate this effect in more detail and to extend the parameter space, micromagnetic simulations \ref{sims} were used. The resulting switching behavior as a function of pulse length and height  for 500\,nm square elements is shown in Fig.\,\ref{fig3}a. For field amplitudes above 22.5\,mT, the simulations show vortex core switching at the leading edge of the pulse, as it was simulated for even shorter pulses by by Hertel \textit{et al.} \cite{Hertel2007}. However, a second region is found where core switching occurs at amplitudes as low as 12.25\,mT, with a strong dependence on the pulse length.  Although it was not identified as such, a similar region can be recognized in the simulations published by Xiao \textit{et al.} \cite{Xiao2007}. In this region, switching occurs just after the trailing edge of the pulse as shown in Figs.\,\ref{fig1} and \ref{fig2}.

We explain this pulse length dependence by the gyration of the vortex around its equilibrium position. In the absence of an external field, this position is located at the center of the element ($P_0$). An applied in-plane field will cause a shift of the equilibrium position to ($P_H$), proportional to the field strength. Assuming a constant gyration frequency, the vortex velocity is proportional to the distance from the equilibrium position. When a rectangular in-plane magnetic field pulse is applied, the vortex starts to gyrate around $P_H$ during the pulse. After the pulse, it gyrates around $P_0$ and the velocity reached in this process is given by its distance to $P_0$. Idealy, the vortex, gyrating around $P_H$, is at a position opposite to $P_0$ at the trailing edge, so that the amplitude of gyration and thus the vortex speed will nearly be doubled (see Fig.\,\ref{fig3} panel e). This defines a coherence condition as shown in Fig.\,\ref{fig3}a, where the critical velocity for core switching \cite{Yamada2007,Guslienko2008} can be reached most easily. This condition is also fulfilled for longer pulses, which end when the vortex is in a similar position after multiple revolutions. As the gyration radius decreases due to damping, the threshold amplitude increases slightly for these longer pulses. 

By the same argument, a pulse ending when the vortex is close to $P_0$ only causes a small-amplitude gyration and the vortex motion is effectively quenched (see Fig.\,\ref{fig3} panel d). This effect was also observed directly in our time resolved imaging (see Movie 3 in the supplementary online material\ref{EPAPS}).  

From our simple analysis, we can derive several properties for this coherent switching scheme: First, the optimum pulse length scales inversely with the gyrotropic resonance frequency of the element, and is thus proportional to the lateral size of the structures for identical thickness\,\cite{Vansteenkiste2008}. The only requirement on rise- and fall times is that they need to be small compared to the respective gyration period. This is confirmed by our simulations: Only little difference is found for values between 5\,ps and 300\,ps for 500\,nm elements. As the losses by damping are limited during the short time before switching, the damping parameter $\alpha$ does not have a significant influence on the threshold amplitudes in the first region: the simulations show less than a 20\% change in switching thresholds for $\alpha$ between 0.005 and 0.05. 
As can be expected, this mechanism does not depend on the exact sample shape. It was confirmed by simulations that disk-shaped samples behave completely analogous.

Experimentally, we confirmed this coherent switching process by a systematic study of several different 500\,nm elements and one 1\,$\mu$m square shaped element with a thickness of 50\,nm. The results are shown in the Fig.\,\ref{fig3}b and \ref{fig3}c. For the 500\,nm elements, the first region was found at pulse lengths between 700 and 900\,ps, with a threshold between 9.5 and 12.6\,mT. This is in perfect agreement with the simulations in Fig \ref{fig3}a, which predict 850\,ps and an amplitude of 12.4\,mT.  For the 1\,$\mu$m element, the experiment yields an optimum pulse length between 1.5 and 1.8\,ns, compared to 1.6\,ns in the simulations. However, in the experiment the minimum pulse amplitude is about 4.6\,mT, which  is 35\% lower than what the simulations predict. 
The asymmetry of the coherent region in the experiment could be attributed to sample heating, which increases with longer pulses. For pulse lengths below 500\,ps,  experimental limitations prevented a detailed investigation.

Evaluating the initial vortex core speed for excitations just below the switching threshold allows to set a lower bound on the critical velocity of the vortex core. For both the 500\,nm and 1\,$\mu$m elements, a value of 250($\pm25$)\,m/s was found, which is in agreement with the theoretical models \cite{Yamada2007,Guslienko2008}. 

In conclusion, time-resolved X-ray microscopy was used to temporally and spatially locate the vortex core switching event triggered by fast monopolar magnetic field pulses. A switching mechanism was found that is based on the coherent excitation of the vortex using both the leading and trailing edge of the pulse. The pulse amplitude and length dependence was investigated by micromagnetic simulations and experimentally verified.  Our analysis shows that the parameters for switching by coherent pulsed excitation do not depend on the details of the complex interaction of a vortex with a time varying magnetic field, but can be derived from only three basic properties of a vortex: The critical velocity, the gyrotropic resonance frequency, and displacement of the vortex structure under static external fields.  Replacing the latter by the displacement for spin transfer torque, these results can be easily applied to excitation by short spin-polarized current pulses. They can also be extended to other schemes using short rectangular pulses, such as bipolar pulses or pulse sequences applied in different directions, by adding up the respective gyration arcs for each pulse segment.
\begin{acknowledgments}
Cooperation with Aleksandar Puzic and Kang Wei Chou is gratefully acknowledged. A.V. acknowledges the financial support by The Institute for the Promotion of Innovation by Science and Technology in Flanders (IWT Flanders). The Advanced Light Source is supported by the Director, Office of Science, Office of Basic Energy Sciences, of the US Department of Energy.
\end{acknowledgments}

\end{document}